\documentclass[aps,prl,twocolumn,letter,longbibliography]{revtex4-1}
\usepackage[english]{babel}

\usepackage{graphicx}
\DeclareGraphicsExtensions{.pdf}

\usepackage{float}
\usepackage{amsmath}
\usepackage{color}

\newcommand{\be}{\begin{equation}}
\newcommand{\ee}{\end{equation}}

\begin{document}
\title{Vibrational properties of hard and soft spheres are unified at jamming}
\author{Francesco Arceri, Eric I. Corwin}
\affiliation{Department of Physics, University of Oregon, Eugene, Oregon 97403, USA}
\date{\today}

\begin{abstract}

The unconventional thermal properties of jammed amorphous solids are directly related to their density of vibrational states. While the vibrational spectrum of jammed soft sphere solids has been fully described, the vibrational spectrum of hard spheres, a model for colloidal glasses, is still unknown due to the difficulty of treating the non-analytic interaction potential. We bypass this difficulty using the recently described effective interaction potential for the free energy of thermal hard spheres. By minimizing this effective free energy we mimic a quench and produce typical configurations of low temperature colloidal glasses. We measure the resulting vibrational spectrum and characterize its evolution towards the jamming point where configurations of hard and soft spheres are trivially unified. For densities approaching jamming from below, we observe low frequency modes which agree with those found in numerical simulations of jammed soft spheres. Our measurements of the vibrational structure demonstrate that the jamming universality extends away from jamming: hard sphere thermal systems below jamming exhibit the same vibrational spectra as thermal and athermal soft sphere systems above the transition. 

\end{abstract}

\maketitle

\textit{Introduction} -- 
Glasses and granular materials are unified by their expression of amorphous rigidity. Seen from the perspective of granular systems, jamming marks the onset of rigidity and occurs at zero pressure, when every particle becomes fully constrained but all contacts are just kissing~\cite{ohern_random_2002-1}. By contrast, in a thermal hard sphere glass, where rigidity is achieved at the dynamical glass transition~\cite{angell_perspective_1988, berthier_theoretical_2011-2}, the jamming point is only reached at infinite pressure when all the particles are forced to come into enduring kissing contact with one another~\cite{charbonneau_glass_2017-3}. As such, the jamming point is a matching point for the two systems, where hard sphere glasses end and soft sphere rigid solids begin. Even though the configurations found in each limiting case must be valid configurations for the other, there is no \textit{a priori} reason to expect that the properties of such configurations should bear any meaningful relation due to their very different origins and interactions. Although the criticality of jamming has been explored from both hard and soft sphere perspectives~\cite{xu_excess_2007, charbonneau_jamming_2015-6, berthier_growing_2016-1}, whether the jamming point represents a smooth crossover between hard and soft spheres or a singular point is still an open question. In this work we demonstrate that for the vibrational properties of both hard and soft sphere systems this point is a smooth joining. We use an effective potential to bring packings of hard spheres to their free energy minima, allowing us to quench towards jamming without the limitations of conventional thermal simulations and to directly measure the vibrational spectrum from the dynamical matrix. 

Amorphous solids exhibit vibrational properties very different from those predicted by Debye theory~\cite{debye_zur_1912-2, zeller_thermal_1971, buchenau_neutron_2007}. The replica mean field theory of glasses and jamming predicts the low-frequency scaling of the vibrational density of states (VDOS) to behave as $D(\omega) \sim \omega^2$ for systems in every spatial dimension~\cite{parisi_mean-field_2010-1, wyart_scaling_2010-2, jacquin_microscopic_2011-1, franz_universal_2015-1}. This non-Debye scaling has been observed numerically in systems of soft spheres right above the jamming point~\cite{charbonneau_universal_2016-5} and is the result of an excess of vibrational modes within this low frequency range. These excess modes are spatially extended but non-phononic and give rise to a peak in the heat capacity of glasses, often called the boson peak~\cite{anderson_anomalous_1971, nakayama_boson_2002, chen_low-frequency_2010}. The VDOS associated with these modes is nearly flat for low frequencies ranging down to a crossover frequency $\omega^\ast$ below which it decays to zero \cite{ohern_jamming_2003-2}. At jamming even an infinitesimal excitation leads to an extended motion and the VDOS is flat until $\omega^\ast = 0$~\cite{yan_variational_2016}. 

In contrast to the mean field picture, as low dimensional soft sphere systems are brought to densities above jamming an additional class of modes appears as quasi-localized modes which are hybridized between system spanning phonons and local rearrangements~\cite{mizuno_continuum_2017}. These modes are believed to control the elastic response to externally applied shears~\cite{maloney_amorphous_2006-1, manning_vibrational_2011-1} and are measured to follow a low frequency scaling of $D_{loc}(\omega) \sim \omega^4$~\cite{mazzacurati_low-frequency_1996-1, xu_anharmonic_2010-1, lerner_statistics_2016, wang_low-frequency_2019}. Such a scaling result has been observed for a wide variety of disordered systems~\cite{silbert_vibrations_2005, baity-jesi_soft_2015-1, widmer-cooper_irreversible_2008-2}. These quasi-localized modes do not appear in the mean field picture as they are exclusively a low-dimensional phenomenon~\cite{morse_two_2019}.

\begin{figure}
\centering
\includegraphics[width=0.95\columnwidth]{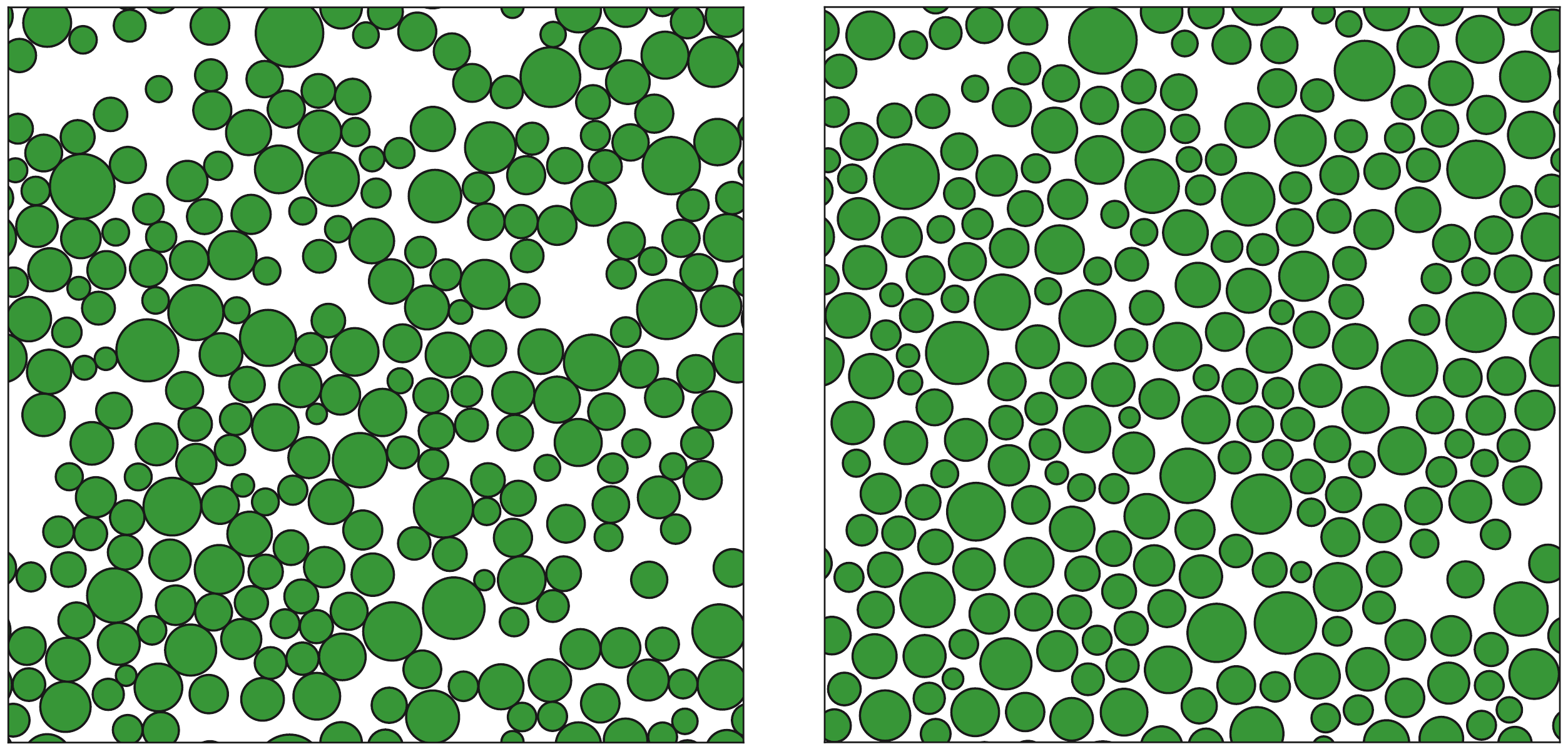}
\caption{Minimization of the effective logarithmic potential in $d=2$ with packing fraction $\varphi = 0.55$. Left: packing after harmonic minimization. Right: final configuration of the same packing after the logarithmic potential minimization.}
\label{fig:minimization}
\end{figure}
%

Similar quasi-localized modes play a central role in the physics of real low-temperature glasses~\cite{lerner_low-energy_2013, ikeda_dynamic_2013-1}. They are described as soft excitations that connect two local minima of the free energy, a scenario introduced by Phillips in the two-level tunneling model~\cite{phillips_low-temp_1971, galperin_theory_1985}. These modes can be derived from anharmonic effects which are directly related to the non-analytic form of the hard sphere potential~\cite{karpov_thermal_1985, buchenau_interaction_1992, gurevich_anharmonicity_2003, ikeda_dynamic_2013-1}. Anharmonic effects independently arise from perturbation theory of hard spheres near jamming~\cite{franz_universal_2015-1, franz_simplest_2016-1}, where the free energy has been found to be well approximated by a logarithmic effective pair potential~\cite{altieri_jamming_2016-3, altieri_higher-order_2018}. The same effective interaction has also been shown in simulations of thermal hard spheres under very high pressure~\cite{brito_rigidity_2006-1} for which an effective medium theory has been developed \cite{degiuli_force_2014}. 

In the limit of high pressure thermal hard spheres this effective logarithmic potential can be understood as deriving from entropic consideration. If the typical timescale between collisions is much smaller than the typical timescale for rearrangements then the time-average of the momenta exchanged between frequently colliding particles is inversely proportional to the gap $h$ between those particles~\cite{brito_rigidity_2006-1}. This coarse graining over time defines a network of effective forces between hard spheres with corresponding potential energy given by a sum of two-body logarithmic potentials of the form 
\begin{equation}
V(h) = -k_BT\log(h).
\end{equation}
Thermal hard spheres near jamming can thus be directly mapped to a collection of athermal particles interacting via the logarithmic effective potential.

While the mean field theory predicts the same vibrational properties for hard spheres below jamming and soft spheres above jamming, in low dimensional systems the vibrational spectra could be very dissimilar due to the very different circumstances giving rise to quasi-localized modes. In this paper, we present a protocol to produce stable glassy configurations based on the minimization of the effective free energy potential for a packing of athermal hard spheres. By measuring the evolution of the vibrational spectrum approaching jamming we show that the spectrum of jammed solids is unified when crossing the transition between the hard and the soft sphere descriptions. This result demonstrates that mechanical and thermal properties of jammed solids arise purely from a geometric origin.

\textit{Numerical methods} -- 
Hard sphere packings compatible with the quench of a thermal glass are produced using the pyCudaPacking package, developed by Corwin~\textit{et al.}~\cite{morse_geometric_2014-5, charbonneau_jamming_2015-6}. The packing is a collection of $N$ particles in $d = 2, 3$ spatial dimensions, with a log normal distribution of particle sizes chosen to avoid crystallization. The packing is inside a box of unit volume with periodic boundary conditions and characterized by the packing fraction $\varphi$, the fraction of the box volume occupied by particles.

Starting from a packing fraction well below jamming we randomly distribute particles and minimize energy using a harmonic interaction potential (the same as used in the context of soft spheres~\cite{durian_foam_1995-1}) to eliminate any overlap between particles. The logarithmic potential is then applied as a pair potential between particles separated by less than a cutoff gap distance. This cutoff is chosen to be twice the value of the position of the first peak of the gap distribution to allow for nearest neighbor interactions and exclude the non-physical next nearest neighbor interactions. However, all the results reported herein are insensitive to this choice as long as the cutoff encompasses nearest neighbors, see Supplementary Materials. We then minimize the potential using the FIRE (Fast Inertial Relaxation Engine) algorithm~\cite{bitzek_structural_2006-2}. 

\begin{figure}
\centering
\includegraphics[width=0.95\columnwidth]{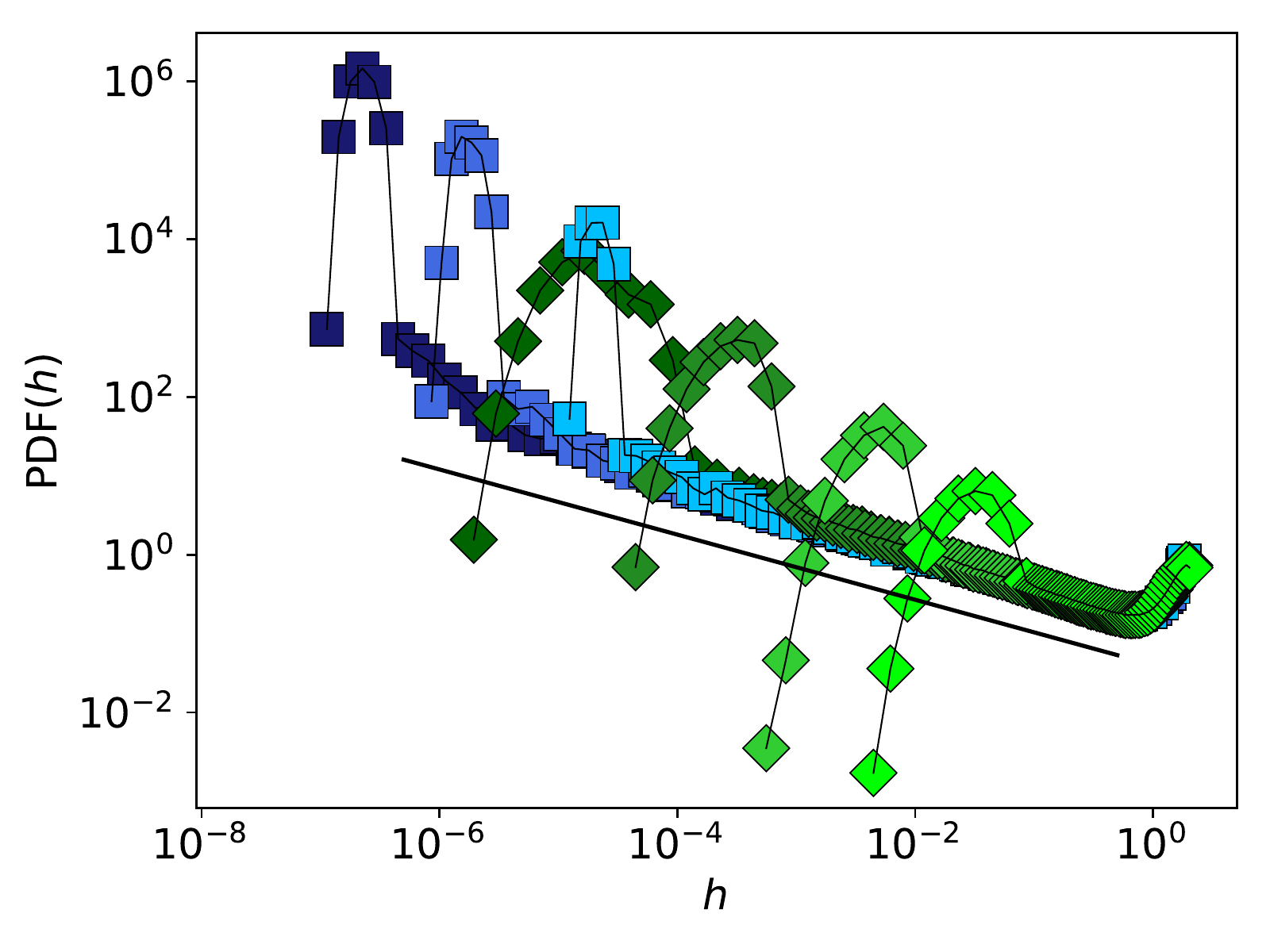}
\caption{Gap distribution of hard sphere packings in $d=3$. The distance from jamming increases from left to right: data from decompressions (blue squares) $\Delta\varphi=1.1\times10^{-7}, 1.3\times10^{-6}, 1.5\times10^{-5}$, data from compressions (green diamonds) $\Delta\varphi=2.3\times10^{-5}, 2.1\times10^{-4}, 3\times10^{-3}, 2\times10^{-2}$. The distributions peak around the value of the typical nearest neighbor gap and then decay following a power law scaling (black line) consistent with the mean field prediction $h^{-\gamma}$ with $\gamma=0.41296...$ \cite{charbonneau_glass_2017-3}. Gaps are cutoff at $h=1$ to avoid showing next nearest neighbor behavior.}
\label{fig:gap}
\end{figure}

The result of the minimization of the logarithmic potential is depicted in Fig.~\ref{fig:minimization}. From an initial packing characterized by a broad distribution of nearest neighbor gaps, the system reaches a configuration where the nearest neighbor gaps are more uniform. This resulting packing is compatible with the time-averaged limit of a thermal hard sphere system, where collisions push particles as far as possible from their neighbors on average. If $\varphi$ is less than the jamming packing fraction $\varphi_J$, no particles are in contact after the minimization and a void region can be found around each particle. We exploit this to creep up in density by inflating particles until saturating 10\% of the minimum gap and then minimizing the effective potential for this new packing fraction. Repeating this procedure iteratively we are able to push the system to a distance from jamming $\Delta \varphi = |\varphi_J - \varphi|$ of the order of $10^{-6}$. To produce packings at densities significantly closer to jamming, we decompress critically jammed soft sphere configurations and then minimize the logarithmic potential~\cite{charbonneau_jamming_2015-6}. By slightly decompressing these packings we maintain the same spatial structure of the jammed systems, with a precise tuning of the distance from jamming $\Delta\varphi$. 

Fig.~\ref{fig:gap} shows the gap distribution from both compressions and decompressions exhibiting the same behavior. We find a power-law scaling of the gap distribution that is well described by the mean field scaling law $\textrm{PDF}(h) \propto h^{-\gamma}$ \cite{charbonneau_glass_2017-3} and has previously been measured for soft spheres precisely at jamming~\cite{charbonneau_jamming_2015-6}. The systems created by decompression from jamming show a sharper peak for the nearest gaps than is found in systems created through compression, even when both systems are nearly the same distance from the jamming transition. This reflects the underlying property that systems created from jammed soft spheres will maintain a memory of their kissing contacts, while those compressed from below have not yet chosen a single set of incipient contacts and thus have a broader distribution.

\textit{Vibrational spectrum analysis} -- 
In order to distinguish extended and localized modes we compute the participation ratio (PR) of each mode, a measure of the fraction of particles that are participating in the motion governed by the mode. Given a mode at frequency $\omega$ with eigenvectors $\{\textbf{u}_i(\omega)\}$, where $\textbf{u}_i$ is the displacement vector for particle $i$, we define the PR as: 
\begin{equation}
\mathrm{PR}(\omega) = \frac{1}{N_s}\frac{(\sum_i^{N_s}|\textbf{u}_i(\omega)|^2)^2}{\sum_i^{N_s}|\textbf{u}_i(\omega)|^4} \; ,
\end{equation}
where $N_s$ is the number of stable particles, i.e. those with at least $z = d+1$ force bearing neighbors \cite{goodrich_finite-size_2012-5}. A mode which corresponds to a totally extensive motion in which every particle participates equally will be characterized by $\mathrm{PR} = 1$, whereas a mode completely localized to a single particle will have $\mathrm{PR} = 1/N_s$~\cite{charbonneau_universal_2016-5, mizuno_continuum_2017}.

\begin{figure}
\raggedright{\footnotesize{(a)}}
\includegraphics[width=0.945\columnwidth]{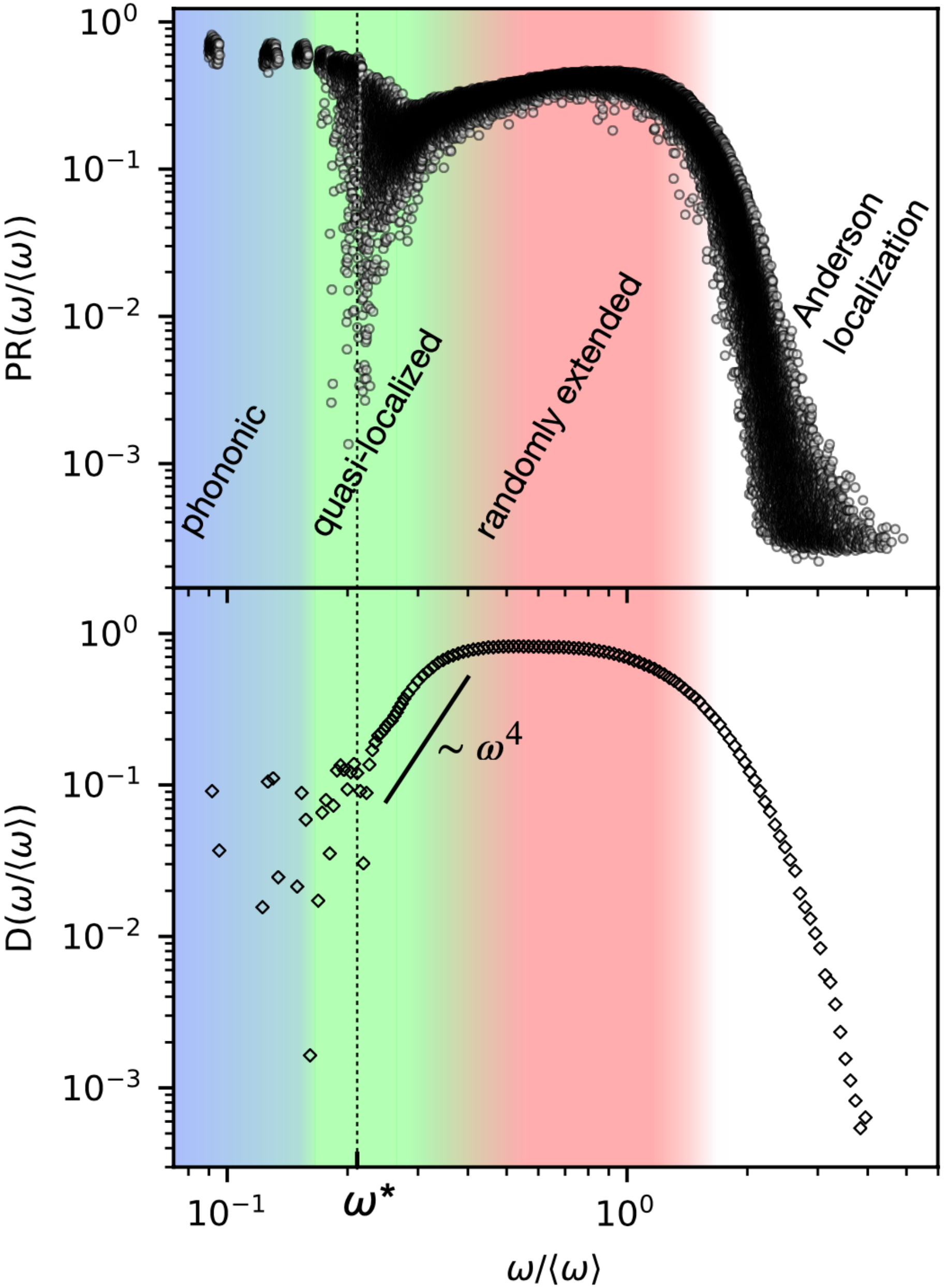}
\raggedright{\footnotesize{(b)}}
\includegraphics[width=\columnwidth]{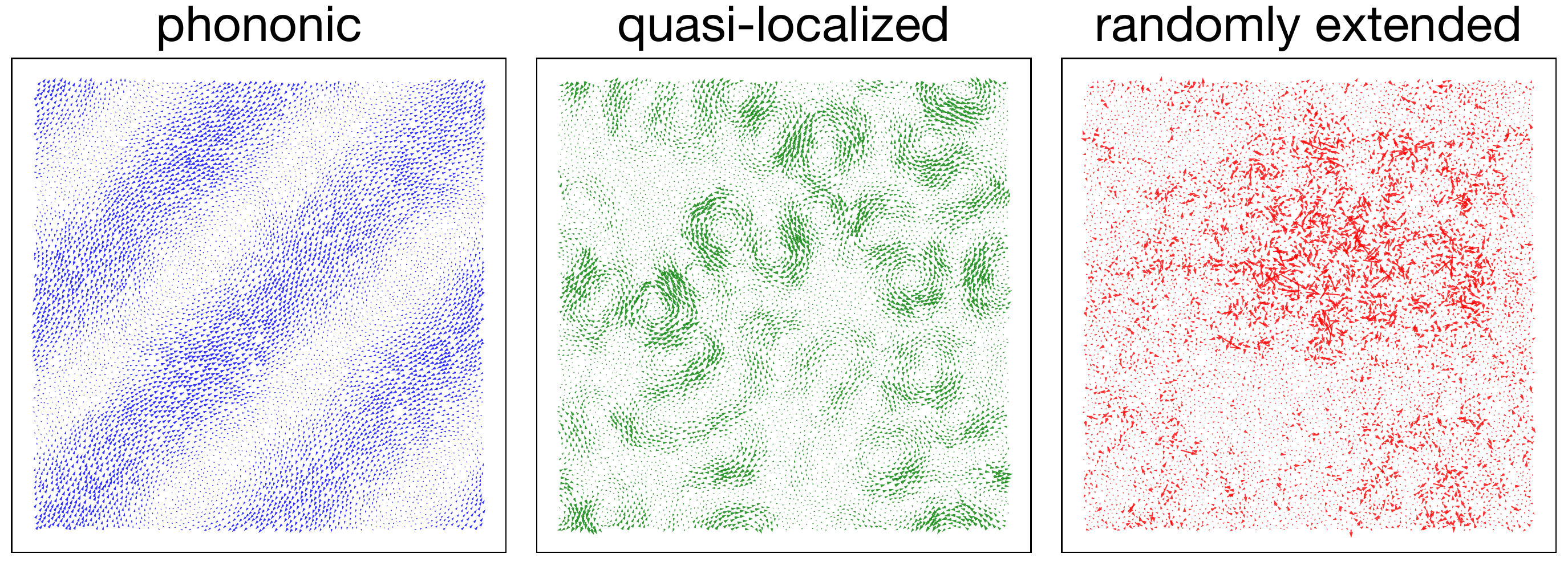}
\caption{a) Participation ratio (PR) and vibrational density of states for a packing of $N=8192$ particles in $d=3$ at $\Delta\varphi = 3\times10^{-2}$. b) Real space representation of the eigenvectors for a packing of $N=8192$ particles in $d=2$ with distance from jamming $\Delta\varphi = 3\times 10^{-2}$. Left: phonon with characteristic plane wave modulation. Center: quasi-localized mode with localized excitations distributed over the whole system. Right: extended anomalous mode which correlates a large portion of the system with random excitations.}
\label{fig:modes} 
\end{figure}

The vibrational spectrum for both two and three dimensional packings produced by minimization of the logarithmic potential can be divided into four different ranges of frequency as illustrated in Fig.~\ref{fig:modes}~a, ranging from lowest to highest frequency: 1) At lowest frequencies, the modes separate into discrete phonon bands with PR $\simeq$ 2/3 as expected for plane waves \cite{mizuno_continuum_2017, wang_low-frequency_2019} (blue region). 2) For frequencies close to $\omega^*$ we find quasi-localized modes which show a splitting in the PR and a power-law decay in the density of states (green region). 3) For higher frequencies modes become increasingly delocalized as indicated by a very high PR. This region corresponds to extended anomalous modes as evidenced by a nearly flat density of states (red region). 4) At highest frequencies modes are strongly localized as a result of Anderson localization in a random medium and have a density of states that decays rapidly with increasing frequency~\cite{xu_anharmonic_2010-1}. 

We analyze the diverse nature of the vibrational modes by looking at the real space representation of their eigenvectors shown in Fig.~\ref{fig:modes}~b. Phonons (left) have a typical plane wave modulation which spans the system. Quasi-localized modes (center), with frequencies near $\omega^\ast$, present a number of localized distortions and vortices hybridized with those phonons at nearby frequencies. Extended anomalous modes (right) contain random seeming excitations spread throughout the entire system.

\begin{figure}
\centering
\includegraphics[width=0.95\columnwidth]{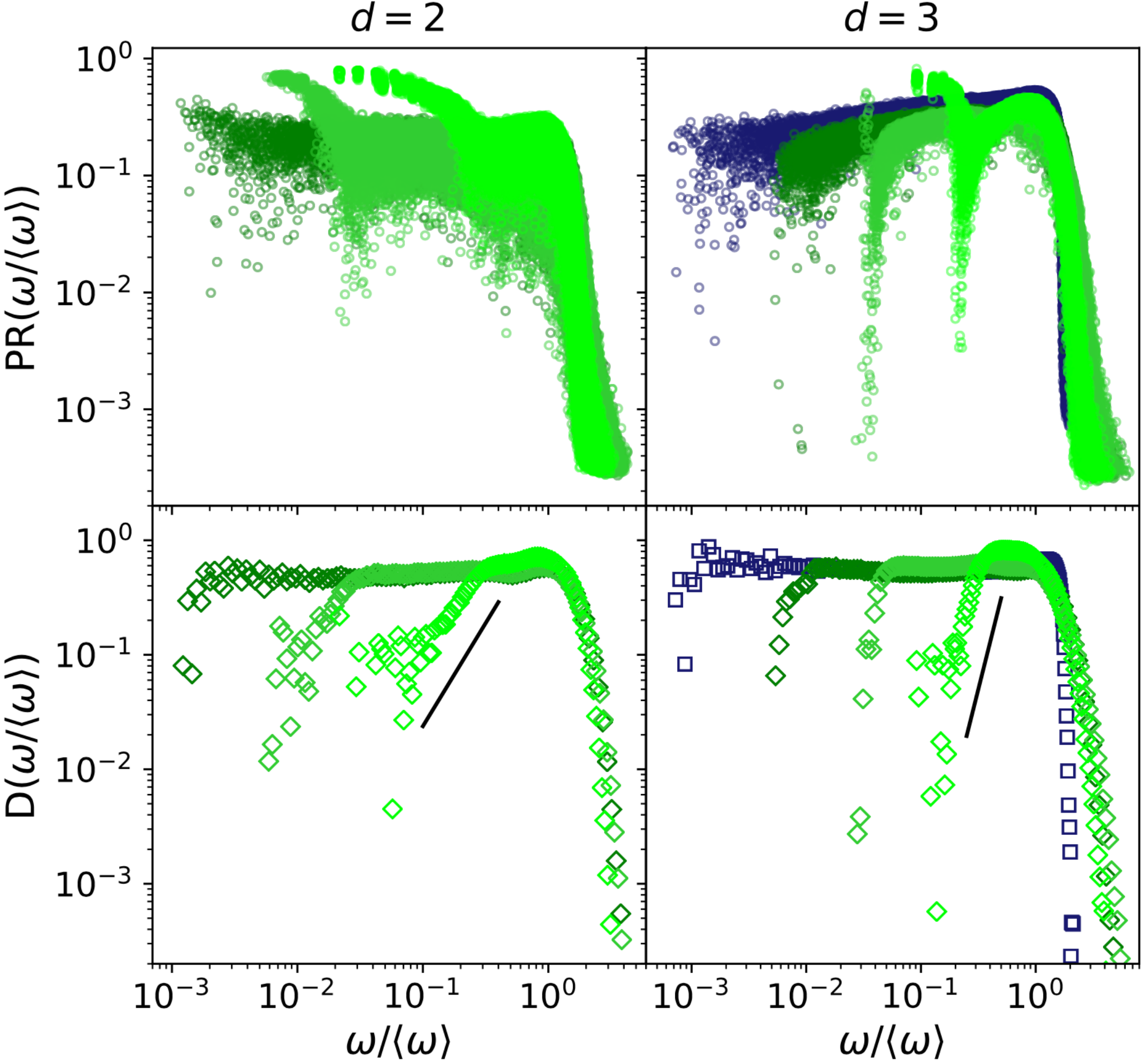}
\caption{Evolution of participation ratio (PR) and vibrational density of states along the compression as a function of $\Delta\varphi$ in $d=2$ (left) and $d=3$ (right). Data from compressions are shown in green and that from decompressions in $d=3$ in blue. Each scatter plot of PR shows data from $10$ samples while the density of state curves are averaged over the same number of samples. The distance from jamming increases from left to right. In $d=2$ $\Delta\varphi=1.1\times10^{-7}$, $\Delta\varphi= 2.3\times10^{-5}, 5\times10^{-4}, 3\times10^{-2}$. In $d=3$ $\Delta\varphi = 2.7\times10^{-6}, 3\times10^{-4}, 3\times10^{-2}$. The low-frequency decay of the density of states in $d=2$ follows $\omega^2$ for every value of $\Delta\varphi$ while in $d=3$ it follows $\omega^4$ sufficiently far from jamming.}

\label{fig:spectrum}
\end{figure}

As shown in Fig.~\ref{fig:spectrum} systems in two and three dimensions differ significantly within the quasi-localized frequency range as evidenced both in the PR and the VDOS. Three dimensional have a greater fraction of modes with strong localization than in two dimensional systems. This difference manifests in the functional form of the decay of the VDOS. For $d=3$ the density of quasi-localized modes dominates over that of extended modes as evidenced by a decay that follows the $\omega^4$ law. For $d=2$ instead, a continuous crossover between phonons and extended modes dominates this region of the spectrum with a decay of the density that goes as~$\omega^2$. These results for hard sphere systems below jamming agree with previous observations for soft spheres above the jamming threshold~\cite{mizuno_continuum_2017}. 

\textit{Criticality near jamming} -- 
Fig.~\ref{fig:spectrum} show the evolution of the density of states and the participation ratio for systems in both $d=2$ and $d=3$ at a broad range of distances from jamming. As jamming is approached quasi-localized modes move toward lower frequencies and hybridize with the existing phonons as local excitations get softer \cite{silbert_vibrations_2005}. For a range of densities sufficiently far from jamming, quasi-localized modes coexist with phonons. For $\Delta\varphi\lesssim10^{-4}$ extended modes dominate the vibrational spectrum. Localized excitations disappear due to the increasing stability of the packing from the compression, a property which translates into a reduction of the number of soft spots from which localized excitations originate \cite{silbert_vibrations_2005}. We observe that for $\Delta \varphi \lesssim 10^{-5}$ localized distortions are suppressed for both spatial dimensions as the extended mode plateau reaches down towards $\omega=0$. We observe that in $d=3$ the low frequency scaling of the VDOS deviates from the $\omega^4$ law while in $d=2$ the $\omega^2$ scaling holds for every step of the compression. 

\begin{figure}
\centering
\includegraphics[width=0.95\columnwidth]{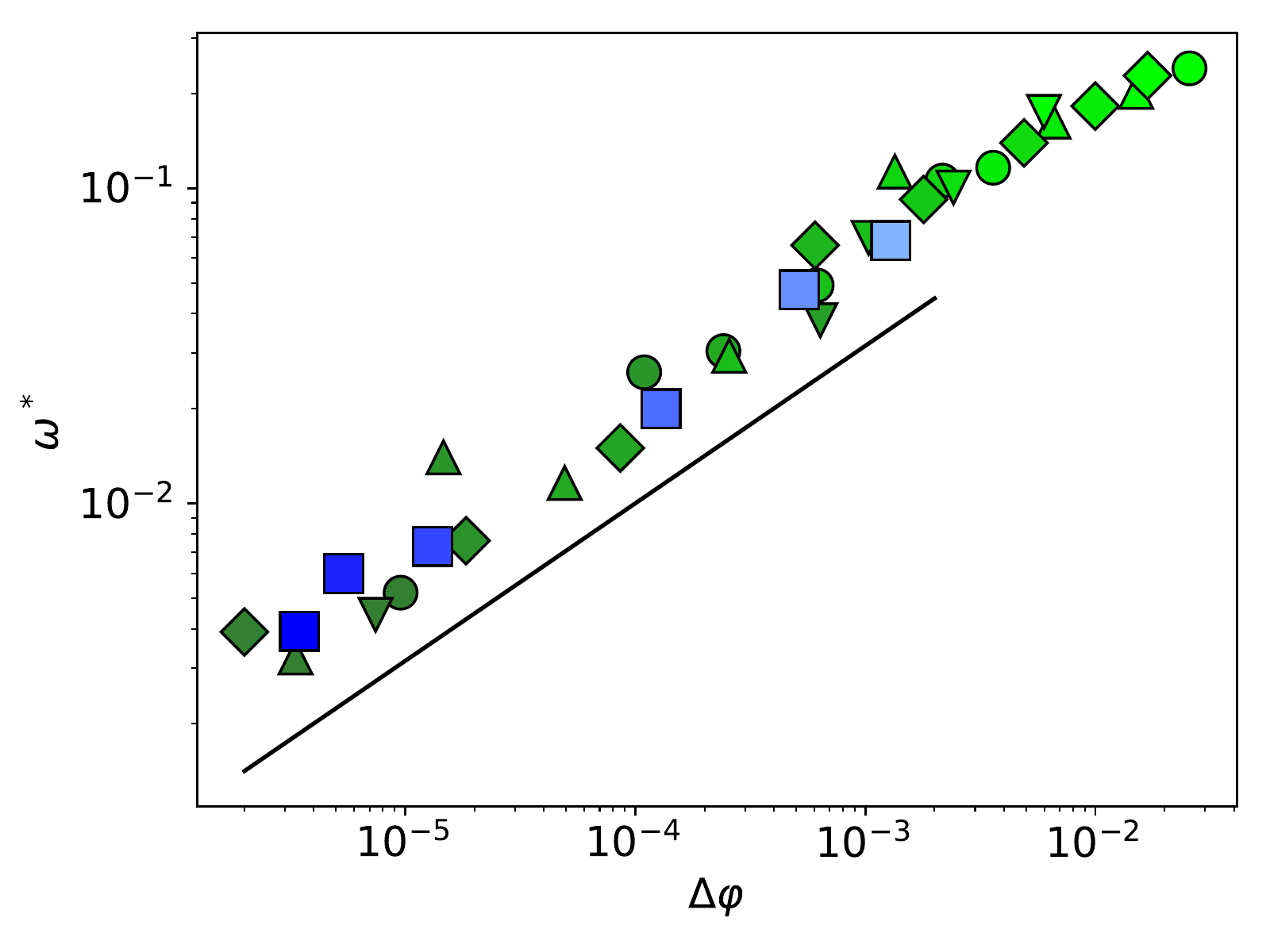}
\caption{Scaling of $\omega^\ast$ as a function of $\Delta\varphi$ for different system sizes from decompressions (blue squares $N=4096$) and compressions (green circles $N=1024$, upward triangles $N=2048$, downward triangles $N=4096$, diamonds $N=8192$). Data are consistent with the critical scaling $\omega^\ast \sim \Delta\varphi^{1/2}$ observed for soft spheres.}
\label{fig:omegastar}
\end{figure}

We measure $\omega^\ast$ as the frequency of the last extended mode above a cutoff in participation ratio, $\mathrm{PR_c} = 8\times10^{-2}$. As shown in the Supplementary Materials, the results are insensitive to the choice of $\mathrm{PR_c}$ for $8\times10^{-2} < \mathrm{PR_c} < 2\times10^{-1}$. The relationship of $\omega^\ast$ on $\Delta\varphi$ is reported in Fig.~\ref{fig:omegastar}. The resulting scaling law is consistent with that already found in the jamming critical region for harmonic soft spheres~\cite{xu_excess_2007}. 

\textit{Conclusions} -- 
By minimizing the logarithmic effective potential we are able to track the structural features from which the mechanical properties of hard sphere glasses originate, both below jamming and at the transition. We have exploited the analytic effective potential to implement a deterministic minimization algorithm and to compute the vibrational properties of hard sphere glasses, something which previously was accessible only from the velocity autocorrelation function in thermal simulations~\cite{ikeda_dynamic_2013-1}. The vibrational modes found below and at jamming using this effective potential quantitatively agree with those observed in soft sphere systems above the transition. Thus, we have demonstrated that granular systems and colloidal glasses have the same vibrational properties sufficiently close to jamming. Further, the scaling of $\omega^*$ confirms that the jamming criticality is universal from both the hard and the soft sides of the transition: thermal hard spheres under very high pressure (or their athermal mapping in this case) have the same criticality as a packing of harmonic soft spheres brought close to zero pressure. 

This work suggests several paths forward for studying hard sphere glassy systems using the tools developed for athermal soft sphere systems. First, it would be useful to apply these techniques to develop a more detailed characterization of the size distribution of soft spots in higher dimension, for which existing methods in identifying quasi-localized modes are not sufficient. Further development of real-space characterizations of these modes will allow for investigations of spatial correlations of quasi-localized modes and how the associated lengthscale evolves towards jamming. Another future direction will be to minimize the logarithmic potential in a previously equilibrated hard sphere glass~\cite{berthier_growing_2016-1}. By doing so, it will be possible to isolate structural features from thermal noise and study mechanical and rheological properties directly related to the real space glassy structure. 

We thank A. Altieri, C. Brito and S. Franz for useful discussions about the logarithmic potential and L. Berthier, E. Flenner, A. Ikeda and A. Liu for fruitful suggestions. This work was funded by the NSF Career Award grant No. DMR-1255370, and the Simons Collaborations on Cracking the Glass Problem (No. 454939 E. Corwin).

\bibliography{LogPotential}

\onecolumngrid

\vspace{10cm}

\centering{\large\textbf{{Supplementary Materials}}}

\vspace{0.5cm}

\begin{figure}[h]
\centering
\includegraphics[width=0.55\columnwidth]{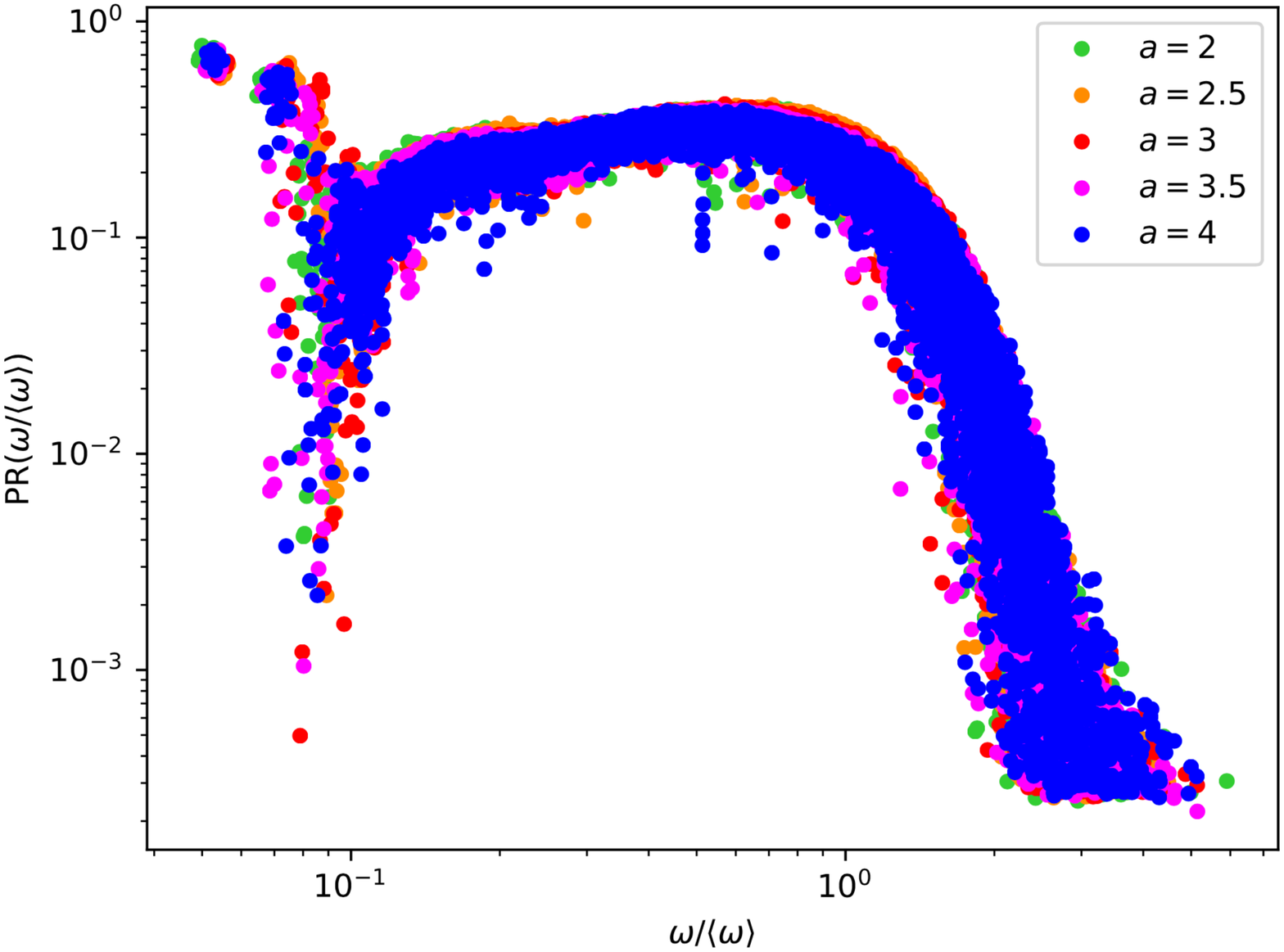}
\caption{Participation ratio (PR) as a function of the gap distance cutoff $h_{cut}$ for the logarithmic potential for a typical sample.  Curves are plotted for $h_{cut} = a h_{peak}$ with $a$ ranging from 2 to 4, where $h_{peak}$ is the size of the gap at the first peak of the gap distribution. Data are obtained by compressing the same initial packing of $N=8192$ particles in $d=3$ from a starting density of $\varphi=0.55$.  The data plotted is for a resulting packing fraction of $\varphi = 0.65722$. The PR does not show any significant difference as the cutoff distance changes over the full frequency range.}
\end{figure}
\vspace{1cm}
\begin{figure}[h]
\centering
\includegraphics[width=0.57\columnwidth]{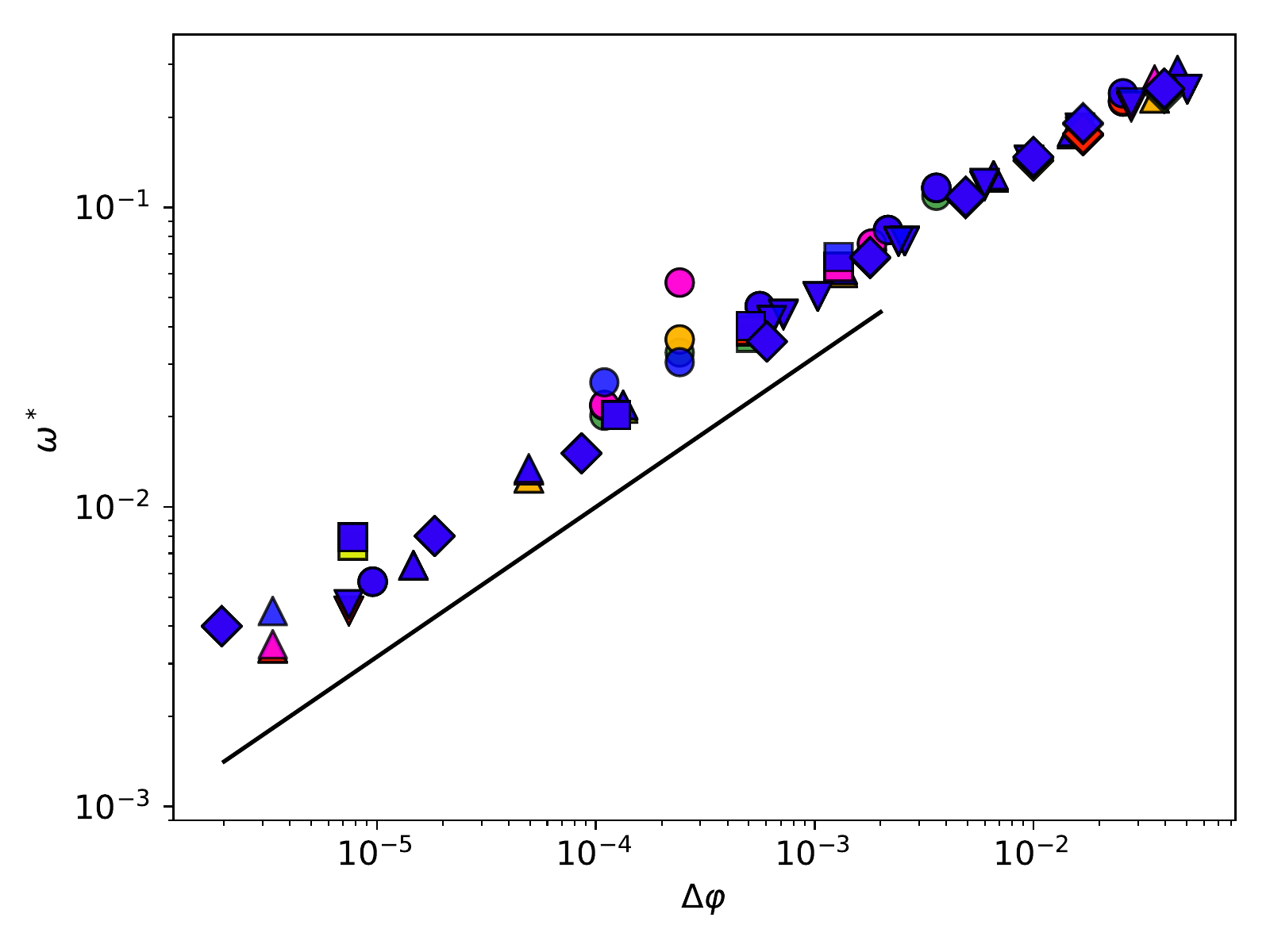}
\caption{Dependence of $\omega^*$ on $\mathrm{PR_c}$, the cutoff threshold for the participation ratio, $\mathrm{PR_c} =$ 0.2 (green), 0.18 (yellow), 0.15 (orange), 0.12 (red), 0.1 (magenta), 0.08 (blue). The curves are plotted for different system sizes from decompressions (squares $N=4096$) and compressions (circles $N=1024$, upward triangles $N=2048$, downward triangles $N=4096$, diamonds $N=8192$). The scaling relation between $\omega^*$ and $\Delta\varphi$ is not affected by the choice of $\mathrm{PR_c}$ for $8\times10^{-2} < \mathrm{PR_c} < 2\times10^{-1}$, values which correspond to $8\%$ and $20\%$ participating particles respectively.}
\end{figure}

\end{document}